# A MACHINE LEARNING ENSEMBLE MODEL FOR THE DETECTION OF CYBERBULLYING

Abulkarim Faraj Alqahtani[1, 2] and Mohammad Ilyas[1]

[1]Department of Electrical Engineering and Computer Science, Florida Atlantic University, Boca Raton, FL, USA
[2]Ministry of National Guard, King Khalid Military Academy, Riyadh 14625, Saudi Arabia

## ABSTRACT

*The pervasive use of social media platforms, such as Facebook, Instagram, and X, has significantly amplified our electronic interconnectedness. Moreover, these platforms are now easily accessible from any location at any given time. However, the increased popularity of social media has also led to cyberbullying.It is imperative to address the need for finding, monitoring, and mitigating cyberbullying posts on social media platforms. Motivated by this necessity, we present this paper to contribute to developing an automated system for detecting binary labels of aggressive tweets.Our study has demonstrated remarkable performance compared to previous experiments on the same dataset. We employed the stacking ensemble machine learning method, utilizing four various feature extraction techniques to optimize performance within the stacking ensemble learning framework. Combining five machine learning algorithms,Decision Trees, Random Forest, Linear Support Vector Classification, Logistic Regression, and K-Nearest Neighbors into an ensemble method, we achieved superior results compared to traditional machine learning classifier models. The stacking classifier achieved a high accuracy rate of 94.00%, outperforming traditional machine learning models and surpassing the results of prior experiments that utilized the same dataset. The outcomes of our experiments showcased an accuracy rate of 0.94% in detection tweets as aggressive or non-aggressive.*

## 1. INTRODUCTION

Today, social networking sites play a significant role in our daily lives. We use social media for various communications, encompassing entertainment, education, personal development, and the workplace. The revolutionary nature of these platforms has made it much easier to connect with people across long distances [1]. Individuals access social media platforms on their cellphones, tablets, and smartwatches, thanks to the widespread and rapid expansion of the internet. Technological advancements have transformed the way we communicate, share information, and interact with communities globally [2]. While social media has many beneficial aspects, it can also be misused. Social media platforms allow users to remain anonymous and conceal their identities, enabling some individuals to abuse these technical capabilities. Bullying, especially cyberbullying, tends to escalate with increased frequency over time. Moreover, the anonymity feature emboldens people to make harsh comments and engage in cyberbullying [3].

The authors in [4] emphasize that cyberbullying profoundly impacts the emotional and psychological well-being of victims. Their study reveals an alarming increase in the frequency of cyberbullying, particularly among teenagers, highlighting it as a significant concern. The authors identify social platforms where cyberbullying occurs, such as X, Facebook, and email. The absence of age restrictions on many social media networks is flagged as a harmful policy with a chilling effect on youths [5]. The misuse of these platforms not only contributes to cyberbullying but also fosters other antisocial acts, rendering them potentially unsafe even for adults. Thus, cyberbullying poses a universal threat, affecting individuals of all ages and locations [6].

Disagreements in viewpoints can escalate into bullying behaviors, defined as aggressive interactions involving words, texts, or tweets between two or more individuals [7]. Those unfamiliar with the benefits of social media may resort to threatening behaviors. The consequences of receiving hateful texts, including anxiety, sadness, self-harm, social and emotional confusion, and even suicidal thoughts and attempts, underscore the gravity of the issue [8]. Consequently, finding a solution to cyberbullying is imperative.

Addressing this issue on social media platforms requires effective mitigation strategies. In these situations, an intelligent and efficient system is necessary. Primary methods for tackling cyberbullying on social media platforms involve Natural Language Processing (NLP), text mining, and machine learning techniques. The challenge lies in the short text length and the presence of syntactic and grammatical errors, making it difficult for NLP algorithms to extract attributes [9]. Trolling, defined by the authors as a concerning online habit involving users with no real-life connection, is directed at specific individuals or groups within the social sphere, often relating to politics [10].

Detecting cyberbullying manually is challenging due to the massive amounts of data; therefore, an automatic detection system can efficiently process large volumes of data using natural language processing, supported by machine learning algorithms. However, identifying cyberbullying text poses a challenge due to variations in language use, leading to inaccuracies in machine learning algorithm classifications, as certain words can alter the context. Consequently, experts worldwide are actively addressing the challenge of accurately determining whether a post or tweet should be classified as cyberbullying.

Despite the introduction of various methods, there remain several issues to address in order to improve the performance of cyberbullying detection. One such challenge is the lack of standardization in preprocessing step sequences, resulting in various adjustments and, subsequently, different levels of accuracy. While feature extractions are commonly utilized for cyberbullying detection, certain features may overlook information about word order. Therefore, it is imperative to explore the effects of alternative strategies for feature extraction. Notably, ensemble models exhibit superior performance compared to traditional machine learning classifiers. In this study, we utilized multiple machine learning classifiers and ensemble models for cyberbullying detection on a Twitter dataset focused on trolling. Our contributions in this study are organized as follows:

- This study aims to develop an approach for cyberbullying detection in both aggressive and non-aggressive tweets. Our approach utilizes several supervised machine learning algorithms and four feature extraction methods.

- We have implemented an ensemble stacking model that effectively combines five well-known algorithms: Decision Tree (DT), Random Forest (RF), Linear Support Vector Classification (LSVC), Logistic Regression (LR), and K-Nearest Neighbors (KNN) using a single stacking model technique.
- The efficiency of cyberbullying detection is assessed based on different feature extraction techniques. We employed well-known methods such as Bag of Words (BoW), Term Frequency-Inverse Document Frequency (TF-IDF), Word2Vec, and GloVe.
- Performance evaluation metrics, including accuracy, precision, recall, and F1 score, are utilized to assess the effectiveness of the selected classifiers. Additionally, our resultsoutperformother approaches that utilizedin the same dataset.
- We compared the results of various feature extraction techniques and applied classification time to determine which features are appropriate for our approach.

The remainder of this paper is structured as follows: 'Literature Review' discusses relevant experiments, 'Methodology' covers dataset description, preprocessing, feature extraction methods, our proposed approach, and machine learning classifiers. 'Results and Discussions' provides a detailed explanation of our experiment's results, and finally, 'Conclusion' summarizes our research and outlines future work.

## 2. LITERATURE REVIEW

Several new methods for identifying cyberbullying have been developed recently. This section provides an overview of various previous studies dedicated to detecting cyberbullying on social media platforms, focusing on both ensemble and traditional machine learning classifiers.

Ensemble models often outperform traditional machine learning models in various domains. In a study by the authors in [11], a stacking ensemble machine learning model was presented to merge two models with varying levels of output performance. SVM and DistilBERT were combined into a single ensemble model using Logistic Regression as a meta-model for classification. Their experiment utilized three datasets related to cyberbullying. For feature extraction, they employed TF-IDF with an SVM model and word embedding with DistilBERT to enhance context. Additionally, they used N-gram for several numbers, and the best results were obtained using unigram, which achieved an accuracy of 85.53%. According to their experiment, the stacking model demonstrated significant results with 89.6% accuracy.

Another research paper [12] proposed an ensemble model that involved five supervised machine learning models and combined these models using a soft voting method. They applied their model technique to classify four different types of binary datasets related to toxic texts, such as hate, offensiveness, and bullying. The ensemble method achieved an estimated 79% accuracy in detecting cyberbullying tweets. When compared to other classifiers used in their experiment, the SVM classifier demonstrated superior performance in identifying cyberbullying.

Moreover, in experiments that utilized ensemble models, the authors in [13] developed ensemble learning models for classifying a cyberbullying dataset that was labeled into two classes: 'offensive' or 'non-offensive' tweets and had around 9093 tweets. They obtained a maximum of 96% accuracy in their experiment. Their approach involved seven machine learning models and combined them into one single model, called a single-level ensemble. They also used a double-level ensemble, which includes four machine learning models in one new model, three models in another novel model, and the integration of these two novel models into one. The double-level ensemble had the best performance in their t performance in their experiment.

While ensemble models were applied in several experiments to detect cyberbullying, some utilized traditional machine-learning models. This illustrates that ensemble learning outperforms traditional machine learning models, enhancing performance and handling large-scale datasets. In this research paper [14], two machine learning algorithms, support vector machine (SVM) and Naïve Bayes, were investigated for training and testing bullying comments and posts on social media platforms. According to their results, both classifiers could identify with 71.25% and 52.70% accuracy. They applied some preprocessing steps that helped improve the performance of the algorithms. They used TF-IDF and polarity to identify the tweets as bullying or non-bullying based on the weighted representation for feature extraction. Additionally, they employed various methods to evaluate their models, such as accuracy, precision, and recall.

Additional research paper[15] employed two machine learning models: Naive Bayes and Logistic Regression to classify live chats of cyberbullying texts. They used around 2000 records for various questions and answers of data that belong to the live conversations of social media. Their result shows that Naive Bayes outperformed Logistic Regression, which achieved 89.79% accuracy in the detection of cyberbullying in live chats.

A few experiments used private datasets, making it challenging to get these detests to try our proposed approach. Thus, we found some experiments using the same dataset we selected which is named Cyber-Troll dataset. The authors in [16] used a deep neural model to detect aggressive tweets. The preprocessing stage is required to clean the data in text classification. Thus, preprocessing includes removing stop words, numbers,punctuation, converting the text into lowercase and Tokenization. After applying preprocessing, they utilized Multilayer Perceptron (MLP), which is a kind of neural network where the nodes or neurons are connected across several layers. Also, they employed designed feature engineering, which means that the authors definite features based on domain knowledge before feeding the model. Feature engineering generates useful input features when developing a machine learning model. Based on their experiment, they chose their features manually to enhance the classification task. They utilized TF–IDF with unigram and bigram encoding for feature extraction, an efficient feature to extract the impotent words that indicate a specific speech. In addition, they utilized another feature extraction that is named word embedding feature. However, TF–IDFoutperforms in their proposed approach. Based on their performance evaluation results which were applied by using four evaluation metrics, their proposed model achieved 92% accuracy and 90 % in precision, recall and F1-score.

In another experiment that used the same dataset we utilized, the authors in [17] employed the Cyber-Troll dataset to detect aggressive or non-aggressive tweets. Also, they added eight emotional features used to classify the tweets based on these emotional features to identify aggressive tweets. Our aim in this paper is to enhance the classification performance, so we quoted their experiment for aggressive classification. They proposed a deep neural network (DNN), trained using Word2vec feature extraction. They started by executing preprocessing steps, including removing unwanted characters, converting text to lowercase, and tokenizing the texts. Based on their statistical results, their proposed DNN model achieved the best evaluation metrics performance. The accuracy of the proposed DNN is 88%, with F1-score, Recall, and Precision in a range of 86% to 87%. In addition, the authors stated that the Word2vec feature is efficient when classifying aggressive tweets in all applied classifiers.

While there are previous experiments in this field, in this paper, we contributed by developing an automatic system for detecting binary classification. Ensemble learning is applied in many classification fields and performs better results [18-19]. Thus, we designed an ensemble method for aggression classification. We used four feature extraction methods to investigate which feature performs better with our proposed approach to know the classification time of each feature extraction. We utilized an ensemble model that combined five machine-learning algorithms using various feature extractions. We also considered our model's accuracy, precision, recall, and F1-score evaluation metrics for detecting aggressive or non-aggressive tweets. Also, we used cross-validation to thoroughly evaluate our model's performance across several folds, guaranteeing accurate and applicable results. Our proposed ensemble method, a well-known stacking model, outperforms the traditional machine learning models regarding aggression classification.

In this study, we propose to enhance the performance of detecting aggressive or non-aggressive tweets using an ensemble learning approach. We aim to compare our results with other experiments using the same dataset. Table 1 compares our approach and other previous experiments for aggression classification.

Table 1. Our scheme compared to recently proposed models.

| Reference | Approach | Dataset | Model |
|---|---|---|---|
| [16] | Deep Learning | Cyber-Troll | Multilayer Perceptron (MLP) |
| [17] | Deep Neural Network | Cyber-Troll | Deep Neural Network (DNN) |
| Our proposed | Ensemble Learning | Cyber-Troll | Stacking |

## 3. METHODOLOGY

In this section, we provide a comprehensive analysis of all the executed machine learning models used in detecting aggressive tweets. Our methodology begins with a description of our ensemble approach and an overview of the dataset employed in this experiment. Additionally, this section covers the preprocessing steps and the evaluation metrics that we utilized to perform our approach.

### 3.1. Ensemble Approach

We utilized supervised models of the dataset (Section 3.2).After applying each supervised model to the dataset, we combined these models into a single ensemble stacking model to improve the overall classification efficiency of the system.The ensemble model is trained by aggregating the predictions of individual models, and the training process involves leveraging the 80% training data to optimize the collaborative learning among the constituent models, ultimately enhancing the ensemble's predictive performance. Thus, this study presents a new method for detecting cyberbullying, focusing on the analysis of aggressive tweets. The significance of automating the identification of offensive text has grown exponentially, emphasizing the need for enhanced performance in this domain. The motivation behind our proposed method to introduce efficient approach of detect binary classification.

To refine our approach, we employed four different feature extraction methods, including Bag of Words (BoW), TF-IDF and various word embedding techniques. This comprehensive investigation aimed to determine the most suitable method for ensemble learning, ensuring a robust and effective cyberbullying detection system. The selection of these feature extraction methods was motivated by the imperative to capture the contextualin the language employed in aggressive tweets.

Before applying these feature extraction techniques, the tweets underwent preprocessing steps. This series of essential steps, including text normalization and tokenization, was undertaken to clean and standardize the textual data. Recognizing the crucial role of preprocessing in improving classifier training and performance, we employed in-depth processing techniques to ensure the clarity of the text.Figure 2 illustrates the workflow of our proposed ensemble approach.

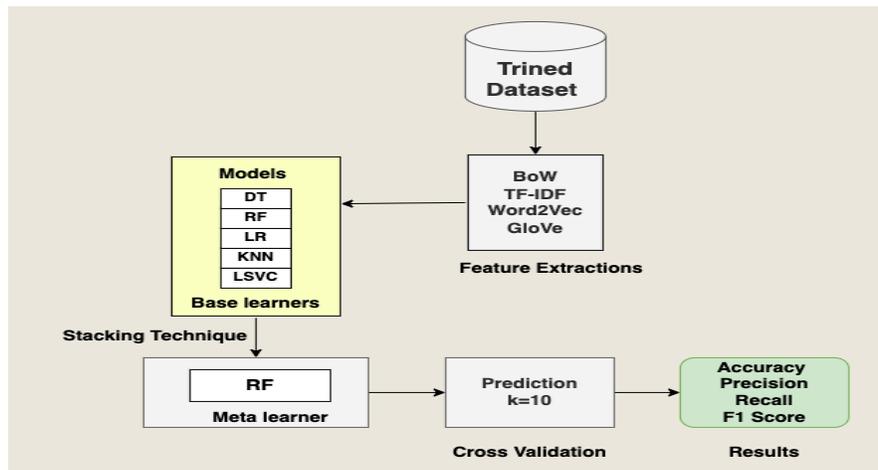

Figure 2.   Illustrates our workflow approach

## 3.2. Evaluating Metrics Calculation

We evaluate our model using accuracy, recall, precision, and F1-score as metrics for assessing average results. Accuracy represents the percentage of correctly classified aggressive and non-aggressive tweets.

$$\text{Accuracy} = (TP + TN)/(\text{TN+TP+FN+FP})$$

Where:

TP = True positive.
TN = True negative.
FP = False positive.
FN= False negative.

When classifying, the total number of aggressive tweets is assigned to the aggressive class, and the total number of non-aggressive tweets is assigned to the non-aggressive class. This is then divided by the total number of predictions [20].

Precision is the proportion of tweets correctly identified as aggressive among all the tweets labeled as aggressive.

$$Precision = (TP)/(TP+FP)$$

The recall measures the proportion of aggressive tweets relative to the total number of tweets in the dataset.

$$Recall= (TP)/(TP+FN)$$

The F1-score is calculated based on the recall and precision of the classifier.

$$F1\ Score = 2*((precision*recall) / (precision+recall))$$

### 3.3. Dataset

The Cyber-Troll dataset, generated by Data Turks [21], serves the purpose of text classification in the English language. This binary-labeled dataset categorizes tweets as aggressive or non-aggressive, aiming to prevent cyber trolls. Labels are denoted by 1 for tweets containing aggressive speech and 0 for those free of aggressive speech. With 20,001 tweets, it includes 12,179 non-cyber-aggressive and 7,822 cyber-aggressive tweets, as illustrated in Table 2.We selected this dataset due to it contains aggressive tweets with keywords intended to hurt and insult others through negative messages. The dataset exhibits an imbalance, with 39% containing aggressive tweets and 61% containingnon-aggressive ones, as appeared in Figure 1.

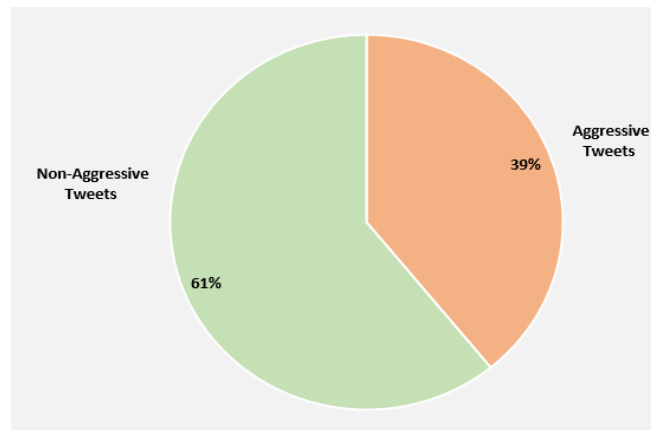

Figure 1. Illustrates a pie chart representing the distribution of classes in the Cyber-Troll dataset

Table 2. Describes the Cyber-Troll dataset.

| Dataset | Total Number of Tweets | Aggressive Tweets | Non-Aggressive Tweets |
|---|---|---|---|
| Cyber_ Troll | 20,001 | 7822 | 12,179 |

## 3.4. Data Preprocessing

Cleaning text can significantly enhance the performance of classification. According to the authors in [22], preprocessing data contributes to achieving better results in natural language processing (NLP) tasks, especially with social media text. Therefore, the raw format of the tweets obtained from the dataset needed adjustments before being fed into the model. The data was cleaned through various preprocessing steps, including removing unnecessary elements (brackets, links, tags, hashtags, white spaces, and numerals), eliminating punctuation, and converting all capitalization to lowercase for uniformity. Once the tweets were cleaned and unwanted elements were removed from the dataset, we performed tokenization to split each single word into tokens. Additionally, for the split dataset, 80% of the train set and 20% of the test set were derived from the cleaned labeled dataset using the Natural Language Toolkit (NLTK). Table 3 displays a sample of the dataset, illustrating the before-and-after results of the preprocessing steps.

Table 3. displays the text both before and after the cleaning process.

| Text before cleaning. | Text after cleaning. |
|---|---|
| What&;s something unique about Ohio? :) | whats something unique ohio |

## 4. FEATURE EXTRACTIONS

The extraction of useful features from the data is a fundamental step in the process, aiding in the proper training of models. Machine learning algorithm efficiency is enhanced by generating secondary features from the original ones [23]. Consequently, the objective of feature engineering is to enhance the accuracy, reliability, and efficiency of algorithm learning by extracting valuable features from raw data [24].

## 4.1. Bag of Words (BoW)

The Bag of Words (BoW) is a popular feature extraction method in NLP, widely used in text classification across various fields due to its simplicity and ease of feature extraction from raw text data [25]. It has proven effective in addressing certain issues in text classification. In our experiment, conducted using the Python language, we imported the CountVectorizer library to apply this feature to all tweets. The strategy of word representation using BoW involves representing each text as a feature vector, with each word treated as a 'token'. A word is represented by 0 if it doesn't appear in the document or 1 if it does.

## 4.2. Term Frequency-Inverse Document Frequency (TF-IDF)

TF-IDF is a feature engineering technique used for extracting features from raw data and is a popular method for statistically evaluating a word's importance in a document [26]. It determines the significance of a word based on its term frequency (TF) and inverse document frequency (IDF). Words with high TF-IDF scores are considered more important.

## 4.3. Word2Vec

Word2Vec is a feature extraction technique used to represent text through word embedding. It generates a unique vector for each word, enabling the identification of connections between words that share meanings and enhancing contextual understanding in the text. The cosine similarity of the vectors indicates the semantic closeness of the represented words [27]. One of the advantages of employing Word2Vec is the reduction in computational complexity it offers, attributed to its pre-training on a large corpus.

## 4.4. Global Vectors for Word Representation (GloVe)

In 2014, the Computer Science department at Stanford University introduced an alternative approach to building word embedding known as GloVe, which relies on unsupervised learning. This method maps words into a space where the distance between them indicates their semantic closeness. GloVe is more complex and challenging to train compared to Word2Vec because it involves computing vectors with each other [28]. The underlying concept of this model is that vectors are assigned numerical values during training based on the semantic information of the words.

# 5. MACHINE LEARNING CLASSIFIERS

The following section outlines the machine learning models employed in this paper. As mentioned earlier, we utilized five ML classifiers for classification, and these models were further combined into one ensemble method.

## 5.1. Decision Tree (DT)

Decision trees belong to the category of supervised machine learning models. Often utilized for solving classification and prediction problems, they are known for their simplicity and capability to extract data features, leading to decisions based on the data. Employing a tree-like structure, decision trees have nodes that represent decisions based on specific features, and branches lead to subsequent nodes. In classification tasks, decision trees predict class labels, while in regression tasks, they make predictions based on the data [29].

## 5.2. Random Forest (RF)

The Random Forest model is constructed based on ensemble learning and trees, which serve as prediction models. This classifier builds a 'forest' of trees, and the strength of the model increases with the number of trees in the forest. This characteristic makes it effective for mitigating overfitting in both classification and regression problems [30]

## 5.3. Linear Support Vector Classification (LSVC)

Linear Support Vector Classificationemploys kernels to tackle prediction challenges, making it a powerful algorithm for solving regression problems [31]. In this study, a linear approach was utilized for binary classification.

## 5.4. Logistic Regression (LR)

LR is a method used for classification by sorting things into groups by finding the best linear and sigmoid maps. Also, it is easy to implement and efficient in binary classification [31].

## 5.5. K-Nearest Neighbors (KNN)

KNN imputes missing data by considering a set of k-nearest neighbors for each sample. For every missing value in a sample, KNN calculates the mean value from its k-closest neighbors in the dataset and uses that average to fill in the missing data for that variable [31].

## 5.6. Stacking Model

A stacking model is a frequently used method for integrating outcomes from various elementary classifiers using a meta-model. It functions as an ensemble learning model designed to aggregate predictions through a voting process [32]. In a stacking model, there are two levels of ensemble methods: the first level comprises individual estimators, and the second level is a combiner estimator of meta-learners that combines all classifiers in the first level to make the final prediction. By utilizing a stacking model, we can enhance predictive capabilities by amalgamating the best features of multiple models across different levels. In practice, we consistently applied the stacking model as two stages: the first stage involved five classifiers (DT, RF, LSVC, KNN, and LR) as base learners, and the second stage used RF in the meta-learner.

# 6. RESULTS AND DISCUSSIONS

Our aim in this research is to detect and classify cyberbullying tweets using an ensemble approach. We employed a stacking ensemble model that incorporates five supervised machine learning classifiers to classify tweets into aggressive or non-aggressive categories. Four different methods for feature extraction, namely BoW, TF-IDF, Word2Vec, and GloVe, were selected to extract features from tweets. We analyze the outcomes of each feature extraction method individually.

## 6.1. Results with BoW Features

We utilized BoW as the feature extraction method for training and testing the classifier. With BoW features, the results demonstrate that Stacking achieves an average F1 score of 0.94%.

Table 4. Displays the results obtained with BoW features.

| Model | Tweets | Precision | Recall | F1-Score | Classification time |
|-------|--------|-----------|--------|----------|---------------------|
| Stacking | 0 | 0.94 | 0.96 | 0.95 | 1min 48s |
| | 1 | 0.94 | 0.91 | 0.92 | |

Table 4 presents the classification evaluation metrics for the Stacking classifiers when using BoW features during both training and testing. Additionally, we conducted a functional classification of time to assess the time consumption for the efficacy of the stacking learning algorithm. As indicated in Table 4, all evaluation metrics yielded positive results ranging from 94% to 95%.

## 6.2. Results with TF-IDF Features

While BoW simply keeps track of term frequencies, TF-IDF goes further by considering the importance of terms, assigning more weight to rare terms. Consequently, TF-IDF enhances contextual comparison compared to BoW, leading to improved performance. Table 2 presents the results of evaluation metrics using TF-IDF for feature extraction.

Table 5.Displays the results obtainedTF-IDF features.

| Model | Tweets | Precision | Recall | F1-Score | Classification time |
|-------|--------|-----------|--------|----------|---------------------|
| Stacking | 0 | 0.95 | 0.98 | 0.97 | 4min 48s |
| | 1 | 0.97 | 0.92 | 0.95 | |

Table 5 displays the evaluation metrics results, indicating improved performance compared to BoW. Precision, recall, and F1 score results demonstrate that the F1 score achieved the best performance in both 0 and 1 classes, signifying excellent matching to the training data. However, it's noteworthy that the classification time is comparatively later than with BoW.

In the previous features, we utilized two features belonging to vector space features. This section and the next one present the results of using features belonging to word embedding techniques.

## 6.3. Results with Word2Vec Features

We utilized Word2Vec features within the context of a stacking model. Table 6 presents the performance evaluation metrics for the stacking model using Word2Vec features. Word2Vec outperformed TF-IDF features, showing a 2% improvement in class 0 and a 3% improvement in class 1 in terms of F1 score.

Table 6.Displays the results obtained withWord2Vec features.

| Model | Tweets | Precision | Recall | F1-Score | Classification time |
|-------|--------|-----------|--------|----------|---------------------|
| Stacking | 0 | 0.96 | 0.98 | 0.97 | 10min 18s |
| | 1 | 0.97 | 0.93 | 0.95 | |

## 6.4. Results with GloVeFeatures

The results of experiments using the GloVe method are presented in Table 7 for the stacking classifier. Training and testing on GloVe features show a performance that is approximately the same as Word2Vec features. Precision has demonstrated better performance in both classes.

Table 7.Displays the results obtained with GloVefeatures.

| Model | Tweets | Precision | Recall | F1-Score | Classification time |
|-------|--------|-----------|--------|----------|---------------------|
| Stacking | 0 | 0.94 | 0.99 | 0.96 | 2min 40s |
| | 1 | 0.98 | 0.90 | 0.94 | |

Cross-validation is a method for assessing the efficacy of data science models. It performs repeated training and testing cycles on different subsets of the data, which gives a more accurate picture of how well the model works than a single train-test split [30]. Therefore, we applied 10-fold cross-validation to perform a thorough assessment of the model's efficacy. Table 8 illustrates the accuracy of the stacking model within four feature extractions.

Table 8. The evaluation results using 10-fold cross-validation of the stacking model with four different feature extraction techniques.

| Features | Accuracy | | | | | | | | | | Mean |
|---|---|---|---|---|---|---|---|---|---|---|---|
| BoW | 0.92 | 0.93 | 0.92 | 0.93 | 0.92 | 0.92 | 0.93 | 0.93 | 0.93 | 0.93 | 92.82 |
| TF-IDF | 0.93 | 0.95 | 0.93 | 0.94 | 0.93 | 0.94 | 0.95 | 0.94 | 0.93 | 0.93 | 94.22 |
| Word2Vec | 0.93 | 0.95 | 0.94 | 0.94 | 0.94 | 0.94 | 0.95 | 0.94 | 0.94 | 0.94 | 93.82 |
| GloVe | 0.93 | 0.93 | 0.92 | 0.94 | 0.94 | 0.94 | 0.94 | 0.92 | 0.92 | 0.93 | 94.00 |
| Model | stacking classifier | | | | | | | | | | |

Based on the results shown in Table 8, the highest accuracy (94.22%) was achieved when using TF-IDF features in the ensemble model approach. The results obtained from four various feature extraction techniques are presented in Tables 4 to 7, representing the evaluation performances for the ensemble stacking model using each feature extraction method separately. In the case of classifying a binary and unbalanced dataset, Word2Vec features yielded the best results in Precision, Recall, and F1-Score. However, the classification time was delayed using Word2Vec. The lowest results were obtained when using BoW, especially for class 1 specifying aggressive tweets, but it had the fastest processing time compared to other feature extraction methods. GloVe obtained better results than BoW and TF-IDF, which was faster than Word2Vec.

In summary, word embedding feature extractions, specifically Word2Vec and GloVe, produced superior results compared to BoW and TF-IDF, except for accuracy, where TF-IDF performed the best. Regarding classification time, BoW was the fastest, taking less than two minutes with the ensemble stacking classifier to complete. However, since our goal is to enhance context, BoW is not the best option for word representation as it represents the vector by 0 or 1. Word embedding features outperformed vector space features, and we consider GloVe features to be the best due to achieving suitable results in all evaluation metrics and being fast in classification time. Thus, GloVe feature extraction is ideal for classification with the ensemble stacking classifier based on three factors: best results, enhanced context, and fast classification time.GloVe attains a 94% accuracy, representing the accurate prediction of outcomes for instances in the dataset. Additionally, it demonstrates high precision 0.96, indicating a low false positive rate, and robust recall 0.95, showcasing effective identification of true positives. The balanced performance is further highlighted by an impressive F1-Score of 0.95.We believe that this model will demonstrate efficiency in addressing cyberbullying, particularly when applied to smaller datasets. Additionally, we anticipate its potential effectiveness on other social media platforms such as YouTube, Instagram, and Facebook.

## 6.5. Comparing Our Method to Other Recent Approaches

We compared our suggested ensemble stacking model approach to two relevant studies, which experimented with four various feature extractions. The study [16] employed the MLP classifier with the Cyber-Troll dataset, utilizing TF-IDF features. A study [17] also applied a DNN model to detect tweets in the Cyber-Troll dataset using Word2Vec features. Table 9 presents a detailed comparison of our approach with these studies.

Table 9. Result of our approach compared to recent approaches utilizing the same dataset.

| Author | Features | Models | Precision | Recall | F1-Score | Accuracy |
|---|---|---|---|---|---|---|
| [16] | TF-IDF | MLP | 90.00 | 90.00 | 90.00 | 92% |
| [17] | Word2vec | DNN | 86.28 | 87.74 | 87.11 | 88% |
| Our Approach | GloVe | Stacking | 0.96 | 0.94 | 0.95 | 94% |

# 7. CONCLUSION

People, especially teens, are increasingly affected by aggressive or cyberbullying comments and postings on social media, which can result in various problems for those who experience it. This paper uses four feature extraction techniques to explore an ensemble stacking classifier for cyberbullying detection. We developed ensemble stacking models to detect aggressive and non-aggressive tweets. We applied five machine learning algorithms separately and combined them into a single stacking classifier. Additionally, four different feature extraction methods, including BoW types and word embedding, were employed in this experiment to investigate the distinctions between these features with our approach. Our ensemble model demonstrated superior performance compared to other experiments. It achieved a 94% accuracy, 96% precision, 95% recall, and 95% F1-Score when utilizing a stacking model with GloVe features, which performed best in terms of results, time, and context. In our comparative study, we contrasted the performance of various approaches. The study establishes the reliability of its findings by comparing them with other experiments using four evaluation metrics. The study's focus on the English language may limit the generalizability of the findings to other languages. Thus, we plan to extend our approach and apply it in different languages, such as Arabic.